\documentclass[useAMS,usenatbib]{mn2e}
\usepackage{times}

\usepackage{graphicx,color}

\input epsf

\usepackage{wrapfig}

\newcommand{\beq}{\begin{equation}}
\newcommand{\beqa}{\begin{eqnarray}}
\newcommand{\eeq}{\end{equation}}
\newcommand{\eeqa}{\end{eqnarray}}

\newcommand{\lsim}{\la}
\newcommand{\gsim}{\ga}

\newcommand{\vect}[1]{\mbox{\boldmath${#1}$}}
\newcommand{\lmk}{\left(}
\newcommand{\rmk}{\right)}

\newcommand{\p}{\partial}

\newcommand{\vex}{{\vect x}}

\newcommand{\vtt}{\vect \tau}
\newcommand{\ven}{\vect n}

\newcommand{\vep}{{\vect p}}

\newcommand{\veX}{{\vect X}}

\title{Resonant Trapping of Stars by Merging Massive Black Hole Binaries}

\author[N. Seto and T. Muto]{Naoki Seto$^{1}$ and Takayuki Muto$^2$
\\
$^{1}$Department of Physics, Kyoto University, 
Kyoto 606-8502, Japan\\
$^2$Department of Earth and Planetary Sciences,
Tokyo Institute of Technology, Tokyo, 152-8551, Japan
}

\begin{document}

\maketitle

\begin{abstract}
A massive black hole binary might resonantly trap a star ({\it e.g.} a white dwarf) and the gas released by its tidal disruption might emit electromagnetic wave signals  around the coalescence of the binary. With post-Newtonian equations of motion including gravitational radiation reaction, we numerically studied  resonant trappings by black hole  binaries with mass ratio 1/100. It is found that 2:1 (and simultaneously 4:2) mean motion resonances of the binaries would be strong  and could, in principle, draw  small third objects deep into relativistic regimes ({\it e.g.} $\sim10$ Schwarzschild radii).  The inclinations of the trapped objects could increase significantly and, in some cases, retrograde orbits could be realized eventually.  
\end{abstract}


\section{Introduction}
Various kinds of mean motion resonances are identified among satellites of Jupiter and Saturn (Murray \& Darmott 2000). Their origins are explained by orbital evolution of the satellites due to their tidal interaction with the planets (Goldreich 1965). As for the orbital motions around the Sun, it is well known that Pluto and Neptune are in a 2:3 resonance (Murray \& Darmott 2000). Mean motion resonances are also found in extrasolar planetary systems (Wright et al. 2011), including a large number of candidates recently provided by  the {\it Kepler} mission (Lissauer et al. 2011). Here, orbital migrations in protoplanetary disks would be the underlying process to yield the resonances (Snellgrove et 
al. 2001; Lee \& Peale 2002, see also Raymond et al. 2008).  Mean motion resonances, once established, are considered to play  fundamental roles on evolution of multiple planet and satellite systems  (see {\it e.g.} Murray \& Darmott 2000).

Orbits of black hole binaries are shrunk by gravitational radiation reaction, and strong gravitational waves from merging massive black hole binaries are promising targets of the Laser Interferometer Space Antenna (LISA, Bender et al. 1998). One of the related phenomena in active debates is potential electromagnetic wave counterparts associated with the mergers. Seto \& Muto (2010) discussed  possibilities that  stars ({\it e.g.} white dwarfs) might be resonantly trapped by massive black hole binaries and electromagnetic wave signals might be produced by the gas released by tidal disruptions of the trapped objects (see  Stone \& Loeb 2010 for similar arguments and also Fujii et al. 2010; Chen et al. 2011 for possible resonant effects not associated with coalescences of binary black holes). In the paper they also examined the coorbital (1:1) resonances numerically and demonstrated that, depending on initial conditions, a test particle can be stably trapped by a black hole binary  around its Lagrangian points $L_4$ or $L_5$ until the epoch close to the coalescence of the parent binary (see also Schnittman 2010; Muto \& Seto 2011).

Stabilities of the coorbital resonances are relatively easy to study both analytically and numerically, but the critical aspect is how to originally settle the trapped objects  around $L_4$ or $L_5$ under the presence of strong gravitational perturbations of the parent black hole binaries. In this paper, we analyze the standard mean motion resonances with ratios of the orbital periods $m:n$ $(m\ne n)$ other than the atypical coorbital ones (with the ratio 1:1).  While there are many parameters to set up the initial conditions for a triple system composed by a   black hole binary and an additional small object,  we here concentrate on the primary question whether a standard mean motion resonance can, in principle, trap a small object stably until a relativistic stage close to the merger of the parent binary. 
%Detailed results would be presented in the forthcoming paper.  
Below, by numerically integrating post-Newtonian (PN) equations of motion for triple systems, we will show preferable results for 2:1 (and simultaneously 4:2) mean motion resonances.

\section{Dynamics of Triple Systems}

\subsection{Mass Ratios}
We study relativistic evolution of triple systems with masses $m_0, m_1$ and $m_2$, including only the gravitational interaction between them.  The particle $m_0(\gg m_1+m_2)$  is the central object and two point masses $m_1$ (inner one) and $m_2$ (outer one) are orbiting around $m_0$. Hereafter we denote $m\equiv m_0$ dropping the subscript 0 for simplicity, and put $m=1$ in the geometrical unit with $G=c=1$. In this subsection, we discuss how to choose the mass ratio $m:m_1:m_2$ for our numerical simulations.

Using  arguments of Hill stability for two particles $m_1$ and $m_2$ orbiting around $m$  in  Newtonian dynamics,  it can be analytically  estimated how closely we can configure two  (coplanar and nearly circular) stable orbits for a given  mass ratio $(m_1+m_2)/m$ (Gladman 1993; Lee et al. 2009). Here we take   $(m_1+m_2)/m=10^{-2}$ roughly corresponding to the stable limit for orbital periods of 2:1. 
For a black hole binary with a larger mass ratio ({\it i.e.} closer to 1), we need to increase the contrast of the two orbital periods  (see also Holman 
\& Wiegert 1999 and references therein) and the involved resonance should be higher than  first order.  
But note that, once a mean motion resonance is realized, it can maintain  stability  for  a higher mass ratio than the simple analytical predictions.  For example,   the allowed mass ratios $(m_1+m_2)/m$ for coorbital (1:1) resonances become 0 according to  Hill stability, but a linear stability analysis around the Lagrangian points $L_4$ and $L_5$ provides the limit $(m_1+m_2)/m<0.04$ (Gascheau 1843).

 For the internal weight between $m_1$ and $m_2$, we  examine the cases with $m_1<m_2$.  This is because the orbital decay due to GW emission is faster for larger mass (see {\it e.g.} Landau \& Lifshitz 1971) and  configuration with relatively approaching  orbits is a key element  for yielding capture into a resonance (Sinclair 1972; Henrard 
\& Lamaitre 1983; Peale 1986). Hereafter we fix $m_1=10^{-5}$ and $m_2=10^{-2}$, considering a fiducial astrophysical  triple system composed by two BHs ($10^5M_\odot+10^3 M_\odot$) and a trapped $1M_\odot$ white dwarf.  The numerical results in the next section are presented for these mass ratios, but we obtained qualitatively similar results for the combination $10^6M_\odot+10^4M_\odot+1M_\odot$.

\subsection{Equations of Motion}
To include general relativistic effects for triple systems, we use the 2.5PN ADM Hamiltonian formally expressed as follows  (Sch{\"a}fer 1987; Jaranowski 
\& Sch{\"a}fer 1997; Galaviz 
\& Bruegmann 2010);
\beq
H=H_N+b_1H_{1PN}+b_2H_{2PN}+b_{2.5}H_{2.5PN},
\eeq
where $H_N$, $H_{1PN}$ and $H_{2PN}$ are the Newtonian, 1PN and 2PN terms. These terms are conservative. $H_{2.5PN}$ is the 2.5PN dissipative term due to gravitational radiation. In eq.(1), we introduced the set of parameters $B_{PN}\equiv (b_1,b_2,b_{2.5})$ for convenience to compare relativistic effects, and our main results are obtained for  $B_{PN}=(1,1,1)$.  In this paper, we negelct spins of the black holes.
\if0%%%%%%%%%%%%%
We also denote $b_{2.5}=1^-$ when we put $b_{2.5}=1$ for $m$ and $m_2$ but at the same  time $b_{2.5}=0$ for $m_1$ to switch off the coherent frictional force induced by radiation reaction for the trapped light particle $m_1$. 
The spins of the black holes are neglected in our study.
\fi%%%%%%%%%%%%%%

The ADM Hamiltonian $H$ is originally  given for the position variables $x_{i\alpha}$ and their conjugate momentums $p_{i\alpha}$ ($i$: label for the particles and $\alpha$: for spatial directions). For example, the Newtonian term is written as
\beq
H_N=\frac12\sum_{i=0}^2    \frac{p_{i}^2}{m_i}-\frac12\sum_{i,j\ne i}^2 \frac{m_i m_j}{d_{ij}}    ,
\eeq
and the 1PN term as
\beqa
H_{1PN}\hspace*{-3mm}
&= &-\sum_{i,j\ne i}^2 \frac{m_i m_j}{4d_{ij}} \Bigg\{ 6\frac{ p_{i}^2}{m_i^2}    -7 \frac{ \vep_{i} \cdot \vep_{j} }{m_i m_j}-\frac{(\ven_{ij}\cdot\vep_i)(\ven_{ij}\cdot\vep_j)}{m_i m_j}\Bigg\}  \nonumber\\
& &-\frac18\sum_{i=0}^2 m_i \lmk \frac{ p_{i}^2}{m_i^2}  \rmk^2
\nonumber\\
& &+ \sum_{i,j\ne i,k\ne i}^2\frac{m_i m_j m_k}{2d_{ij}d_{ik}},
\eeqa
where we defined $p_i^2\equiv \vep_i\cdot\vep_i$, $d_{ij}=|\vex_i-\vex_j|$ and $\ven_{ij}=(\vex_i-\vex_j)/d_{ij}$. 
The second and third summations in eq.(3) represent interactions between different particles.
Meanwhile, the 2PN term $H_{2PN}$ is composed by many  elements, and we use its lengthy expression given in  Lousto and Nakano (2008) (see   Appendix A in their paper).  The 2.5PN term $H_{2.5PN}$ can be found {\it e.g.} in eq.(41) in  Jaranowski \& Sch{\"a}fer (1997).

In this paper, we introduce   new variables $s_{i\alpha}\equiv p_{i\alpha}/m_{i}$ to  improve accuracy at numerically handling triple systems with  large mass ratios (Seto \& Muto 2010; Muto \& Seto 2011). This prescription  is  based on the technical reason, and we fully use the symmetrical three body Hamiltonian without introducing approximations associated with $m_3\ll 1$ (except for the numerical experiments discussed in Appendix A). In other words, the three particles are dealt equivalently in the post-Newtonian framework.  Our equations of motion are given by appropriate partial derivatives as
\beq
\frac{\p x_{i\alpha}}{\p t}=\frac1{m_i}\frac{\p H}{\p s_{i\alpha}}, ~~  \frac{\p s_{i\alpha}}{\p t}=-\frac1{m_i}\frac{\p H}{\p x_{i\alpha}}.\label{eom}
\eeq
The right-hand side of these equations are well behaved even in the test particle limit $m_i\to 0$. The 1PN contributions to the equations of motion  can be derived from  eq.(3), but their  explicit forms are also found   in  Lousto and Nakano (2008) (see their eqs.(4) and (5)). For the  partial derivatives of the 2PN Hamiltonian, we generate their fortran forms using {\it Mathematica}.  The dissipative 2.5PN contributions for eq.(4) are presented {\it e.g.}   in Galaviz 
\& Bruegmann (2010) (their eqs.(12) and (13)).  For our triple systems, the total number of equations becomes 18.
	They are integrated numerically as described in the next section.

\subsection{Orbital Parameters}

We denote  the coordinate distances $d_1\equiv |{\bf x}_1-{\bf x}_0|$, $d_2\equiv |{\bf x}_2-{\bf x}_0|$ and $d_{12}\equiv |{\bf x}_1-{\bf x}_2|$ between three particles $m_1, m_2$ and $m(=m_0)$. The semimajor axes $a_i$ and the eccentricities $e_i$ of the inner particle ($i=1$) and the outer particle ($i=2$) are numerically determined from the consecutive maximum (apocenter distance $a_i(1+e_i)$)  and the minimum (pericenter distance $a_i(1-e_i)$) of the radial separation $d_i$ between the central object $m$ and the particle $m_i$.  The orbital angular parameters are defined in a standard manner as shown in Fig.1 (see also Murray 
\& Dermott 2000).
The basic variables for studying resonant states are the longitude of ascending node $\Omega_i$, the argument of pericenter $\omega_i$ and the mean anomaly $M_i=2\pi(t-\tau_i)/T_i$ ($\tau_i$: time of the pericenter passage, $T_i$: the interval between the passages). In the case  of Newtonian elliptical motion, $M_i$ is related to the true anomaly $f_i$ (shown in Fig.1) as 
\beq
M_i=f_i-2e_i \sin f_i+\frac34 e_i^2 \sin 2f_i+{\cal O}(e_i^3).
\eeq

%%%%%%%%%%%%%%%%%%%%%%%%%%%%%%%%%%%%%%%%%%%%%%%%%%%%%%%%
\begin{figure}
%\epsscale{1.3}
%\plotone{1.eps}
\includegraphics[width=110mm]{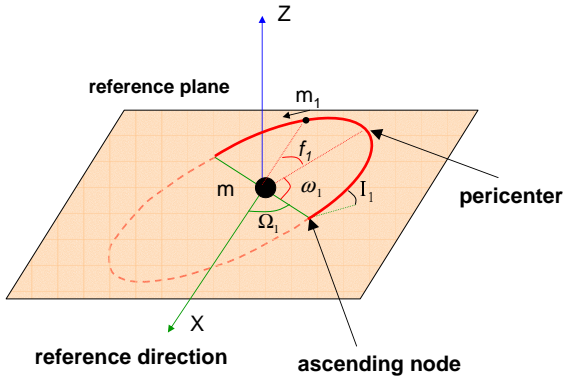}
\caption{ Definitions of orbital angular parameters. We take the reference plane (normal to the Z-axis) as the initial orbital plane of the outer particle $m_2$. The reference direction on the plane is the $X$-axis. The (green colored) objects are on the reference plane, and the (red colored) ones are on the (precessing) orbital plane of the inner particle $m_1$.   The longitude of the pericenter  is defined by $\varpi_1=\Omega_1+\omega_1$ with the longitude of ascending node $\Omega_1$ and the argument of pericenter $\omega_1$ shown above. The mean longitude is written by $\lambda_1=M_1+\varpi_1 $ with $M_1=f_1+{\cal O}(e_1)$.  The parameters $\varpi_2$ and $\Omega_2$  for the outer particle  $m_2$ are defined similarly (for $\lambda_2$ see \S 3.3).  During the evolution of the system,  the particle $m_2$ nearly stays on the reference plane with $I_2\sim 0$.
}
\label{fig1}
\end{figure}
%%%%%%%%%%%%%%%%%%%%%%%%%%%%%%%%%%%%%%%%%%%%%%%%%%%%%%%%%%%

\section{Numerical Analyses}

\subsection{Initial and Termination Conditions}
We set up the initial conditions of our numerical calculations in the following manner. The outer particle $m_2(=10^{-2})$ is placed at $d_2=300$ with $e_2\simeq0$. The inner particle $m_1(=10^{-5})$ is released around the pericenter at the distance $d_1=300 q^{2/3}$ with a control parameter $q$ ($q\in [0.27,0.99]$ at 0.01 interval).  We set its initial eccentricity $e_1\sim 0.15$ and the relative inclination $I_1\sim 0.01$ [rad]. Although we do not go into details about the earlier histories of our triple systems, we expect that such an orbital alignment would be realized {\it e.g.} through interaction with the disk around the central black hole $m$.
But, as we see later, exact coplanarity is not a critical requirement for yielding a stable resonant trapping. 
From Kepler's law and the relation $d_i=a_i (1-e_i)$ for the pericenter distance, the ratio of the initial  orbital periods $T_1$ and $T_2$ is given as $T_1/T_2\sim (a_1/a_2)^{3/2}\sim 1.2q$.

We define the reference plane for the orbital angles  by the initial orbital plane of the outer particle $m_2$ (see Fig.1). 
 Given $m_1\ll m_2$, the total initial angular momentum of the triple is nearly normal to the reference plane (parallel to the $Z$-axis in Fig.1). The two particles $m_1$ and $m_2$ are initially set nearly at conjunction with the argument of pericenter $\omega_1\sim\pi/2$.  

Next we describe when we stop our runs. Given our perturbative (post-Newtonian) treatment of the relativistic effects, we conservatively terminate the integration when one of the following close encounters happens for the first time; (I) $d_1<10m$, (II) $d_{12}<10m_2$ and (III) $d_2<10m$. We set  another termination condition (IV) $d_1>30d_2$, as a simple ejection of the lightest particle $m_1$ from the system. These criterions are selected  primitively for particle systems. But the tidal radius of a white dwarf (mass $1M_\odot$ and radius $10^4$km) is given by $\sim10^5  (m_2/10^3M_\odot)^{1/3}$km for a black hole of mass $m_2$, and larger than $10m_2\sim 10^4(m_2/10^3M_\odot)$km used for  the condition II. Therefore, the white dwarf would be tidally disrupted before the condition II is applied.  On the other  hand, the central black hole $m$ can directly swallow a white dwarf.

We solve the equations of motion (4)
 using a fifth order Runge-Kutta scheme with an adaptive step size control (Press et al. 1996).
Here we briefly describe two of the  numerical tests done for our code.
These are performed for the specific sets of parameters  $(a_2,e_2)=(100,0.1)$ and (50,0.1).

(i) Binary evolution by gravitational radiation reaction.  For a  purely binary (tentatively putting $m_1=0$), we compare our  numerical results $da_2/dt$ and $de_2/dt$  with the corresponding analytical expressions obtained  by Peters (1964).  Since the latters are given for $B_{PN}=(0,0,1)$ in our post-Newtonian parameterization,  we prepared the numerical results in the same setting $B_{PN}=(0,0,1)$.  We found that the differences between the analytical and numerical ones  are less than 1\%.

(ii)  Conservation of  Hamiltonian for triple systems.  For  several resonant triple systems  with $m_1\ne 0$ and  $B_{PN}=(1,1,0)$ (namely excluding the dissipative 2.5PN terms), we  checked  conservation of the total Hamiltonian $H=H_N+H_{1PN}+H_{2PN}$. The variation rate  $d H/dt$ is less than $10^{-5}$ of the expected energy loss rate due to the 2.5PN terms.

\subsection{Final States}

Our primary interest in this paper is stable  resonant evolution of  relativistic triple systems.  Therefore, we first provide the final semimajor axes $a_{2,fin}$ of the outer particles  $m_2$ at the end of our calculations (started from $a_{2,ini}=300$). In Fig.2,  the results $a_{2,fin}$  are shown with the filled symbols.  These are obtained with the parameters $B_{PN}=(1,1,1)$.  The squares are for the case I (close encounter between the inner one $m_1$ and the central object $m$), the circles for the case II (close encounter between $m_1$ and $m_2$), and triangles for the case IV (ejection of $m_1$).  No run ended with the condition III. Fig.2 shows that, for $q=0.33\sim0.38$, the inner particles $m_1$ were stably trapped by the outer ones $m_2$ down to relativistic regime  $a_{2,fin}\sim 20$. These are due to 2:1 mean motion resonances.  In the next section,  we pick up one of them ($q=0.33$) and follow its evolution.  Another dip around $q=0.55$ was caused by 3:2 resonances that exist from the beginnings of the runs (as expected from Hill stability).  

\if0%%%%%%%%%%%%%%%%%%%%%%%%%%%%%%%%%%
During these resonant evolutions, we have the relative angle $|\Omega_1-\Omega_2|\sim \pi$, reflecting that the direction of the total angular momentum is almost unchanged. We will return to this issue later by following  evolution of a specific run. 
\fi%%%%%%%%%%%%%%%%%%%%%%%%%%%%%%%%%%%%%%

In Fig.2 we also show the inclination $\sin I_1$ around $\sim 500$ cycles  before the end of our calculations. The crosses are for retrograde orbits with $|I_1|>\pi/2$ and the pluses are for prograde ones with $|I_1|<\pi/2$.  While we always had $I_2\sim0$ for the outer particle, the angle $I_1$ could grow significantly. In particular, for deeply trapped ones (with small $a_{2,fin}$), the inner orbits could even become retrograde $|I_1|>\pi/2$. Many of them ended with the condition I.  But the condition II is realized in some cases ({\it e.g.} $q=0.40$ with $a_{2,fin}\sim 60$), and, for the fiducial system containing a white dwarf, we might observe electromagnetic wave signals associated with the tidal disruptions.  One the other hand, in our numerical samples, the ejections of $m_1$ were common outcomes for the 3:2 resonances.

In addition to the standard setting $B_{PN}=(1,1,1)$ mentioned so far, we  examined  evolution of the triple systems with $B_{PN}=(1,0,1)$ and $(0,0,1)$ to study relativistic effects. In this comparison, the three particles are again handled equivalently without using approximations based on  $m_1\ll 1$, and  the dissipative 2.5PN term was remained to generate orbital migrations.  The results are presented in Figs.3 and 4. We found that the overall behavior of Fig.2 is unchanged for the 1PN case given  in Fig.3 with $B_{PN}=(1,0,1)$.  The minimum value of the final semimajor axis $a_{2,fin}\sim20$ is similar to the 2PN results, and retrograde orbits are also observed.  In contrast,  the Newtonian results in Fig.4 with $B_{PN}=(0,0,1)$ are quite different from other two.
For example, we no longer have retrograde inner orbits.  While the 2:1 resonances are realized at lower $q$, compared with the 2PN and 1PN cases,   the minimum value of the final semimajor axes increases to  $a_{2,fin}\sim 75$ in Fig.4.  These results indicate that the relativistic effects help the long term resonant trapping of binary black holes.

\if0%%%%%%%%%%%%%%%%%%%%%
As mentioned earlier, the lighter particle $m_1$ should be injected on a inner orbit to realize a resonant capture by  gravitational radiation. For confirmation, we quickly investigated the opposite initial states with $1<q<2.5$, and did not find a capture event. At $q\gsim 2$, the outer light particle $m_1$ is left behind the inspiral of the inner particle $m_2$.
\fi%%%%%%%%%%%%%%%%%%%%%%

%%%%%%%%%%%%%%%%%%%%%%%%%%%%%%%%%%%%%%%%%
\begin{figure}
%\epsscale{0.95}
%\plotone{rin.eps}
\includegraphics[width=80mm]{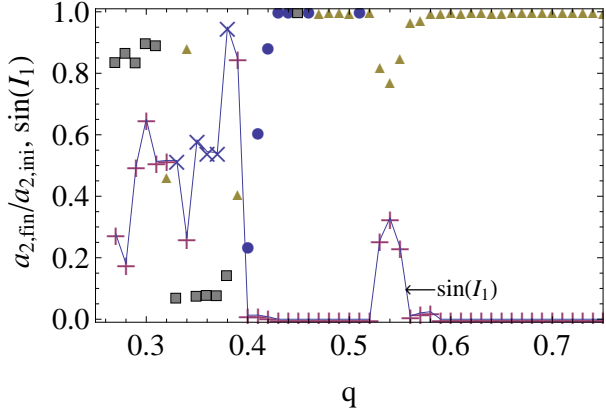}
\caption{ The final states of the triple system with masses $m=1$ (central object), $m_1=10^{-5}$ (inner particle) and $m_2=10^{-2}$ (outer particle). The results are obtained with the full 2.5PN equations of motion [$(b_1,b_2,b_{2.5})=(1,1,1)$]. The initial semimajor axis of $m_2$ is $a_{2,ini}=300$ and the ratio of initial two orbital periods is $T_1/T_2\sim1.2q$.   The filled symbols represent the final  
semimajor axes  $a_{2,fin}$     [squares for  $d_1\le 10m$ (case I),  circles for $d_{12}\le 10m_1$ (case II) and   triangles for ejection with $d_1>30d_2$ (case IV)].  The crosses and pluses on the solid curve show the inclination $\sin I_1$  at $\sim 500$ rotation cycles before the termination of our integrations (crosses: retrograde orbits with $|I_1|>\pi/2$, pluses: prograde with $|I_1|<\pi/2$). 
}
\label{fig2}
\end{figure}
%%%%%%%%%%%%%%%%%%%%%%%%%%%%%%%%%%%%%%%

%%%%%%%%%%%%%%%%%%%%%%%%%%%%%%%%%%%%%%%%%
\begin{figure}
%\epsscale{0.95}
%\plotone{rin.eps}
\includegraphics[width=80mm]{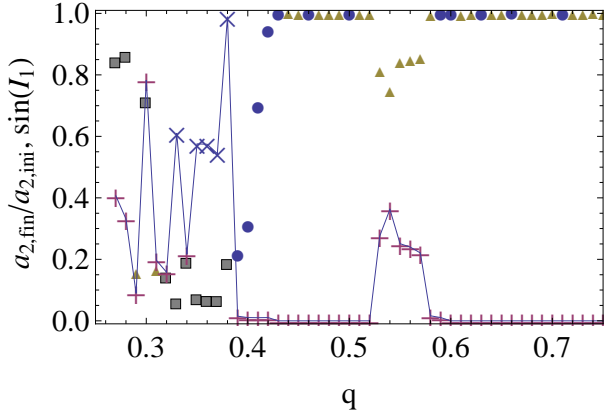}
\caption{ Same as Fig.2 but the 2PN term is  excluded [namely with $(b_1,b_2,b_{2.5})=(1,0,1)$]. The overall profiles of the semimajor axes and the inclinations are similar to the results in Fig.2 obtained with 2PN terms. 
}
\label{fig3}
\end{figure}
%%%%%%%%%%%%%%%%%%%%%%%%%%%%%%%%%%%%%%%

%%%%%%%%%%%%%%%%%%%%%%%%%%%%%%%%%%%%%%%%%
\begin{figure}
%\epsscale{0.95}
%\plotone{rin.eps}
\includegraphics[width=80mm]{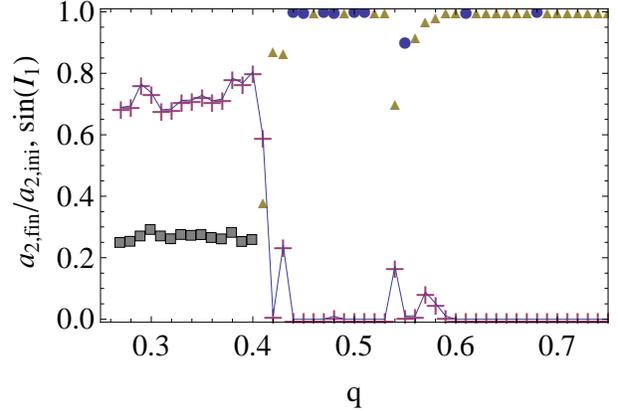}
\caption{ Same as Fig.2 but the 1PN and 2PN terms are excluded [namely with $(b_1,b_2,b_{2.5})=(0,0,1)$].  We no longer have retrograde orbits (crosses appeared in Figs.2 and 3) and minimum value of the final semimajor axes increased to $a_{2,fin}\sim 75$. 
}
\label{fig4}
\end{figure}
%%%%%%%%%%%%%%%%%%%%%%%%%%%%%%%%%%%%%%%

\subsection{Evolution of Orbital Parameters}

Here we pick up the full 2.5PN run for the initial parameter $q=0.33$, and examine its time evolution. Our integration ended at $a_{2,fin}=22$ where the retrograde inner particle $m_1$ has a close encounter  with the central one (the termination condition I).  This run is a typical example for the long-term 2:1 resonances appeared in Fig.2 around $q=0.35$   with the filled squares.  
For reference, we also show the corresponding Newtonian result (with $q=0.33$) in Fig.6

Since the orbital decay is determined by gravitational radiation and we have the simple relation $m\gg m_2\gg m_1$, the semimajor axis $a_2$ is  related to the remaining time $t_r$ and the rotation cycles $N_r$ before the coalescence of $m$ and $m_2$ as (Peters 1964)
\beq
t_r\simeq\frac5{256} \frac{m^2}{m_2} \lmk \frac{a_2}m\rmk^4,~~N_r\simeq \frac{m}{64\pi m_2}\lmk \frac{a_2}m\rmk^{5/2}
\eeq
 (or $t_r\sim 3.1(a_2/100m)^4$yr and $N_r\sim 5.0\times 10^4(a_2/100m)^{5/2}$ for the fiducial triple with $m=10^5M_\odot$ and $m_2=10^3 M_\odot$). Therefore we use the semimajor axis $a_2$ of the outer particle to describe the evolutionary stages of the system. The basic results are summarized in Figs.5 and 6 with time elapsing leftward from  $a_2=300$.  The PN effects largely change the evolution of  the eccentricity and the inclination of the inner particle. Below, we mainly comment on the PN calculations given in Fig.5.

We show the semimajor axis $a_1$ of the inner orbit in Fig.5a. It initially oscillates until the resonant capture around $a_2\sim 285$ (as confirmed later through a resonant argument), and then smoothly shrinks as $a_1\sim 2^{-2/3} a_2$.
In Figs.5b and 5c, we show the eccentricities $e_1$ and $e_2$ of the two particles. They show short period oscillations. The inner one becomes $e_1\sim 0.2$ around $a_2=250$, while the outer one keeps small values $e_2<2\times 10^{-3}$. 
%, and, even with the large mass ratio $m_2/m_1=10^3$, 
%evolution if the outer one $m_2$ is affected by the inner one.

The relative angle $\Omega_2-\Omega_1$ is presented in Fig.5d. It soon settles to $\Omega_2-\Omega_1\sim \pi$. 
This means that the components of the angular momentums of $m_1$ and $m_2$ projected to the reference plane are nearly in the opposite directions.
The oscillation in the early stage (with small $|I_2|$), would be partly due to our choice of the reference plane.  For the outer particle, we initially have $I_2=0$ and the longitude of the ascending node $\Omega_2$ is not well defined.  But other angles $\lambda_1, \lambda_2, \Omega_1,\varpi_1$ and $\varpi_2$ can be determined almost unaffected by this choice, as understood from their intrinsic geometrical meanings. 
Since we have $|I_2|\ll 1$ and $e_2\ll 1$, we determine the mean longitude $\lambda_2$ using the time when $m_2$ pass the $XZ$-plane in Fig.1.  
With $m_1\ll m_2$, the initial total angular momentum is nearly normal to the reference plane.

In Fig.5e, we show the inclination angle $I_1$. The magnitude of the angle $I_2$ for the outer one stays at $I_2\sim 0$, as expected from $m_1\ll m_2$.  At the Newtonian order, the angular momentum $J_i$ of the particle $m_i$   is related to the orbital period $T_i$ as $J_i\propto m_i T_i^{1/3} (1-e_i^2)^{1/2}$. Therefore,   we define the combination 
\beq
\Sigma\equiv \sqrt{1-e_1^2}\sin I_1-1000\cdot 2^{1/3}\sqrt{1-e_2^2} \sin I_2 \label{sigma}
\eeq
 for the component of the total angular momentum projected to the reference plane, and plot it in Fig.5e.   
The result $\Sigma\sim 0$ together with the relation $\Omega_2-\Omega_1\sim \pi$ shows that the orientation of  the total angular momentum is nearly  parallel to the $Z$-axis and almost unchanged. A similar behavior is observed in the Newtonian results (see Fig.6e) and also in numerical simulations of resonant planet migrations (Thommes 
\& Lissauer 2003). These adiabatic results are reasonable, since the time scale of the precessions  ${\dot \Omega}_i$ is much shorter than that of the orbital decay.

In Figs.5f and 5g, we present the resonant arguments
\beq
\theta_{e1}=2\lambda_2-\lambda_1-\varpi_1,~~~\theta_{I1}=4\lambda_2-2\lambda_1-2\Omega_1
\eeq
defined for a 2:1 eccentricity resonance and a 4:2 inclination resonance respectively. Due to the symmetry with respect to the reference plane, the inclination resonances start from the second order in the standard classification of resonances (Murray \& Darmott 2000). As shown in panel Fig.5f, the argument $\theta_{e1}$ initially circulates, but turns to librate around $\theta_{e1}\sim0$ at $a_2\sim 285$ where the inner particle is captured into the 2:1 eccentricity resonance. The libration amplitude gradually decreases, and the inclination $I_1$ shortly increases up to $\sin I_1\sim0.2$.  On the other hand, throughout this run, the relative angle $\varpi_2-\varpi_1$ and thus the resonant argument $\theta_{e2}=2\lambda_2-\lambda_1-\varpi_2$ circulates (see Fig.5h).  But for some other initial parameters $q$ ({\it e.g.} $q=0.36$ with the final values $(a_{2,fin},I_1)$ similar to $q=0.33$, see Fig.2), we found structured distribution for the combination $\varpi_2-\varpi_1$.  Our Newtonian run in Fig.6h also shows the libration $\varpi_2-\varpi_1\sim \pi$ at $a_2\lsim 280$.

%%%%%%%%%%%%%%%%%%%%%%%%%%%%%%%%%%%%%%%
\begin{figure}
\includegraphics[width=85mm]{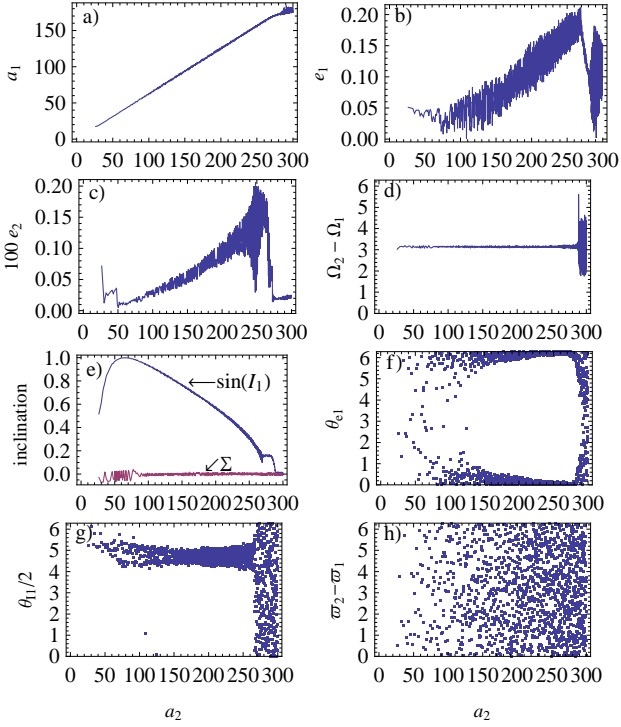}
%\epsscale{1.}
%\plotone{rsam.eps}
\caption{ Evolution of the triple system with $q=0.33$ and $B_{PN}=(1,1,1)$. The horizontal axis is the semimajor axis $a_2$ of the outer particle from the initial value $a_{2,ini}=300$ down to the final value $a_{2,fin}=22$. The calculation is terminated by the condition II  (close encounter between $m_1$ and $m_2$). The panel a) shows the semimajor axis of the inner particle.   b) and c)  The eccentricities $e_i$  of the inner and outer particles  respectively.  d) The relative angle $\Omega_2-\Omega_1$ between the longitudes of ascending nodes. e) The inclination $\sin I_1$ and the combination $\Sigma$ defined in Eq.(\ref{sigma}) for the total angular momentum. The inner orbit becomes retrograde at $a_2\lsim 60$.  f), g) The resonant arguments  $\theta_{e1}=2\lambda_2-\lambda_1-\varpi_1$ and $\theta_{I1}/2=2\lambda_2-\lambda_1-\Omega_1$ plotted once every 500 rotations of $m_2$. h) The relative angle $\varpi_2-\varpi_1$
}
\label{fig5}
\end{figure}
%%%%%%%%%%%%%%%%%%%%%%%%%%%%%%%%%%%%%%%%

%%%%%%%%%%%%%%%%%%%%%%%%%%%%%%%%%%%%%%%
\begin{figure}
\includegraphics[width=85mm]{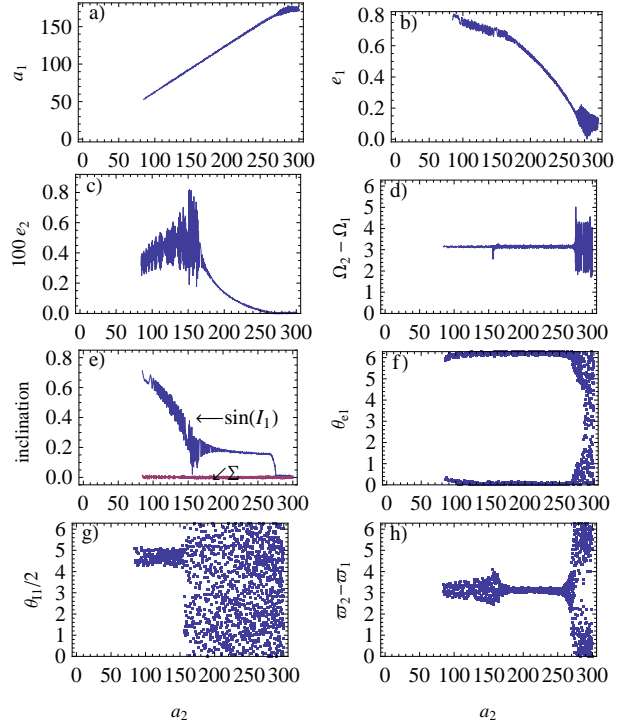}
%\epsscale{1.}
%\plotone{rsam.eps}
\caption{ Same as Fig.5 but without the 1PN and 2PN terms ($B_{PN}=(0,0,1)$). This run  ended with the condition I around $a_2\sim 80$.
}
\label{fig5}
\end{figure}
%%%%%%%%%%%%%%%%%%%%%%%%%%%%%%%%%%%%%%%%

At $a_2\sim 265$, in addition to $\theta_{e1}$, the argument $\theta_{I1}$ is captured into libration around $\theta_{I1}/2\sim3\pi/2$.  From Figs.5d, 5f and 5g, we can see  that other two arguments, $4\lambda_2-2\lambda_1-2\Omega_2$ and $4\lambda_2-2\lambda_1-\Omega_1-\Omega_2$ are also in libration states, and we have $|\varpi_1-\Omega_1|\simeq \pi/2$ as in the case of the Kozai libration (Kozai 1962) that is caused by secular terms and known to be vulnerable to relativistic effects in the course of orbital migration. 
In Fig.5g, the  distinct outlier point  around $a_2=110$ is associated with $e_1\ll 1$ for which the mean longitude $\lambda_1$ would not be well-behaved. 
In the Solar system, two Saturnian satellites, Mimas and Tethys are in the inclination resonance with the argument $4\lambda_2-2\lambda_1-\Omega_1-\Omega_2$.

 Concurrently with the libration of $\theta_{I1}$, the inclination $I_1$ starts to increase due to accumulation of resonant effect. Similar evolution appeared  in the 3:2 resonances. Growth of inclination accompanied by simultaneous (eccentricity and inclination) resonances is  observed also in numerical simulations of  planet migrations in Newtonian dynamics (Thommes 
\& Lissauer 2003). But, there, relatively large eccentricities $e_1\sim 0.6$ are required to yield the simultaneous resonances. Interestingly, our Newtonian simulation in Fig.6  shows that we indeed need  large eccentricities $e_1\sim 0.6$ for  the onset of the  simultaneous resonances around $a_2\sim 160$.

Now we briefly discuss the differences between the PN and Newtonian calculations. Here, it is important to note that the simultaneous resonances ($\theta_{e1}$ and $\theta_{I1}$, see eq.(8)) are realized with the relation ${\dot \Omega}_1\sim {\dot \varpi}_1$.  First we examine the Newtonian run given in Fig.6.
In the  eccentricity resonance (with the argument $\theta_{e1}$) for a nearly coplanar Newtonian system with  $e_1\ll 1$, the inner pericenter  retrogrades as  $0>{\dot \varpi}_1\propto 1/e_1$  (Murray 
\& Dermott 2000).  In Fig.6, the simultaneous resonances ($\theta_{e1}$ and $\theta_{I1}$) are established   at $a_2\sim 160$ when the eccentricity $e_1$ reaches 0.6. In the earlier stage $a_2\gsim 160$, we found $0> {\dot \Omega}_1> {\dot \varpi}_1$, reflecting the smaller eccentricity $e_1$ and the resultant higher regression rate ${\dot \varpi}_1$. 
Thus the condition ${\dot \Omega}_1\sim {\dot \varpi}_1$ is first met at the relatively large eccentricity $e_1\sim0.6$. 

Next we analyze the PN calculations given in Fig.5.   The PN effects generate a positive contibution  $({\dot \varpi}_1)_{PN}>0$ to the apsidal precession (Landau \& Lifshitz 1971).  Meanwhile, from close comparisons between the runs in Figs.5 and 6,  we found  that the nodal precession rate ${\dot \Omega}_1$  is less affected by the relativistic effects for the nearly coplaner triples. Therefore, the difference $ {\dot \varpi}_1-{\dot \Omega}_1<0$ for the Newtonian results at $e_1\lsim 0.6$ is  compensated by the PN correction $({\dot \varpi}_1)_{PN}>0$, and the condition ${\dot \varpi}_1\sim {\dot \Omega}_1$  (thereby increase of the inclination $I_1$) can be realized at a smaller $e_1$, as seen in  Fig.5.

\if0%%%%%%%%%%%%%%%%%%%%%%%%%%%%%%%%%%
  These indicate  the importance of relativistic apsidal precessions to make ${\dot \varpi}_1\sim {\dot \Omega}_1$ from a   smaller eccentricity $e_1$ and thereby increase the  inclination $I_1$.  
The large inclination angle $I_1$ widens the minimum separation between $m_1$ and $m_2$ and would contribute to maintain their mutual stability, as expected from Fig.2.
\fi%%%%%%%%%%%%%%%%%%%%%%%%%%%%%%%%%%%%%%%%

During the stable resonant evolutions of the two cases, close encounters between $m_1$ and $m_2$ are avoided.  The inner eccentricity $e_1$ shows remarkable contrast between Fig.5b and Fig.6b, and this is an important parameter for the condition I, as we see below.
 In Fig.5,  the inner particle is on a polar orbit relative to the outer one around $a_2\sim60$, and then retrogrades with the final value $\sin I_1\sim 0.5$ ($I_1\sim 150^\circ$).  Just before  the end of this run, the eccentricity $e_1$ goes up (not sufficiently captured in Fig.5b), and the calculation is ended by the condition I with the pericenter distance  $(1-e_1)a_1\sim 10$. In contrast, for the Newtonian sytem in Fig.6b, the inner eccentricity $e_1$   increases almost monotonically, and the condition I is met earlier at $a_2\sim 75$.

\section{Discussions}
We have studied relativistic evolution of the resonant triple system with the post-Newtonian approach including terms up to 2.5PN order. We mostly fixed masses of the system at $m=1$ (central object), $m_1=10^{-5}$ (inner particle) and $m_2=10^{-2}$ (outer particle). Here we briefly summarize our basic results. We found that  an  eccentricity-type  2:1  resonance (together with inclination-type 4:2 resonances) is strong and  the outer particle could, in principle,  trap the inner one down to the relativistic regime with final distance $a_2\sim 20$.  
Here, the PN effects play important roles for the resonant evolution. For example, they help the onset of the simultaneous resonances at a small eccentricity $e_1$.  For the PN system, the inner eccentricity $e_1$ can remain small, and the close encounter between $m$ and $m_1$ is delayed, compared with the corresponding Newtonian simulation in which the eccentricity $e_1$ continues to grow. 
During the simultaneous  resonances, we observed  rising of the mutual inclination angle, and even retrograde orbits could be realized. 

% Here relativistic effects play  important roles and the
% inner orbit could eventually retrograde.

Next we describe astrophysical implications of our results,  particularly for gravitational wave astronomy. As an example, we follow the evolution presented in Fig.5 and introduce relevant physical units by taking $m=10^5M_\odot$. The resonant trapping of the inner one $m_1$ continues down to  $a_2\sim 22$,  where the  system becomes unstable and a close encounter between $m_1$ and $m$ occurs with distance $d_{1}\sim 10m$. At this point, the gravitational wave frequency from the main binary $m$-$m_2$ is $2/T_2\sim 6$mHz. The time before its coalescence is $t_r\sim 2.7$ days and we have $2N_r=2300$ gravitational wave cycles left.
%(0.6mHz, 27 days and 2300 cycles respectively  for $m=10^6M_\odot$). 
Before this transition, the inspiral wave of the main binary has a phase correction of ${\cal O}(m_1/m_2)$ due to the trapped particle $m_1$, compared with a purely two body system $m$-$m_2$.  It might be possible to effectively describe the waveform of the triple system in  a simple manner with a small number of additional fitting parameters. Otherwise we might not make a long term coherent integration of the gravitational wave signal before the transition.  

If the trapped particle $m_1$ is a neutron star (or a stellar mass black hole) and plunges into the outer black hole (as for $q=0.40$ in Fig.2), the associated gravitational wave signal might be detected by future ground-based interferometers as a precursor to the merger of the parent binary that might be observed with space interferometers. But these compact objects would also have chances to fly out from the systems at very high velocities $\gsim 0.1c$. 
In the case of a white dwarf, we might observe  characteristic flares of electromagnetic waves due to the gas released by its tidal disruption (Rees 1988).  Detections of such electromagnetic wave signals would have significant impacts on astronomy and cosmology ({\it e.g.} identification of host galaxies).

\if0%%%%%%%%%%%%%%%%%%%
In this paper, we have reported the primary  aspects of the relativistic resonant evolution
 of triple systems.
\fi%%%%%%%%%%%%%%%%%%
 Here, we comment on possible expansion of our study.
While we have mainly investigated triple systems already captured in first-order resonances, it would be meaningful to analyze the earlier evolutionary stages. The later stages around the binary coalescence should be also studied, including strong gravity beyond our perturbative treatment. Spins of black holes might become important there, especially for highly inclined orbits. In addition, various astrophysical processes ({\it e.g.} interaction with a  gas disk, tidal deformation of extended stars) would be worth studied.

This work was supported by JSPS grants  20740151 and 22.2942.

%\input{ref3}

%\begin{thebibliography}{10}

%\end{thebibliography}

\appendix

\section{Coherent radiation reaction force on the small body}

When two particles $m_1$ and $m_2$ move incoherently around $m$, the frictional forces caused by gravitational radiation can be regarded as independent. However, in the case of resonant motions, there can be coherent frictional effect between them.  For an eccentric inner particle in a 2:1 resonance, we can expect this in the quadrupole order of gravitational radiation, as indicated by coupling between quadrupole moments of the two particles. Using  the approximate formula for radiation reaction force given in Landau \& Lifshitz (1971), we roughly estimated the frictional work imposed on the inner particle $m_1$ by the quadrupole  moment of the outer one $m_2(\gg m_1$), and found that this work alone would not be sufficient to keep the resonant state with respect to the infalling outer one. 

 For reference, we also made the following numerical  experiment in order to shut down the coherent frictional force caused by  the  main binary on the small inner particle. 
In eqs.(4), we  put $B_{2.5PN}=0$ for the lightest body ($i=1$) but keep $B_{2.5PN}=1$ for other two bodies ($i=0$ and 2). Then we obtained numerical results similar to Fig.2. 

The standard resonant transfer of energy and angular momentum from $m_1$ to $m_2$ would be the dominant and sufficient process to keep the stable resonant evolution as seen in Fig.2. In the case of  a coplanar triple with $e_2\ll 1$, such transfer would be realized by 
   conjunctions  after the  pericenter passages of the inner particle (see {\it e.g.} Murray 
\& Dermott 2000).  The pericenter are known to retrograde in Newtonian dynamics, and its rate is proportional to $\propto e_1^{-1}$ (Murray 
\& Dermott 2000).

\end{document}